\documentclass[showpacs,preprintnumbers,amsmath,twocolumn,amssymb]{revtex4}

\usepackage[dvips]{graphics,graphicx}
\usepackage{subfigure} 

\usepackage{amsfonts}

\begin{document}


\title{RF Shimming Pulses For Ex-Situ NMR Spectroscopy and \\ Imaging
 Using B1 Inhomogeneities}

\author{Louis-S. Bouchard}
 \affiliation{Materials Sciences Division, Lawrence Berkeley National
 Laboratory and Department of Chemistry, University of California, 1
 Cyclotron Rd, Berkeley CA 94720 }

\date{February 14, 2006}

\begin{abstract}
I describe a method for generating ``shim pulses''  for NMR
spectroscopy and imaging (MRI) by taking advantage of the inherent
inhomogeneity in the static and radiofrequency (RF) fields of a
one-sided NMR system. The RF
inhomogeneity here is assumed, without loss of generality, to be a
linear gradient. General polynomials in the spatial variables can be
generated using $x$, $y$ and $z$ RF gradients using trains of hard
pulses which result in linear combinations of monomials $xy$, $y^2$,
$xz$ etc., and any desired scalings of these monomials. The basic shim
pulse is constructed using small tip angle approximations.
\end{abstract}



\maketitle

\newfont{\Bb}{msbm10}

\newcommand{\dotprod}{{\scriptscriptstyle \stackrel{\bullet}{{}}}}

\section{Introduction}

``Shim pulses'' have been proposed by Topgaard and
Pines~\cite{bib:topgaard_shimpulses} using pairs of
adiabatic pulses in combination with gradient modulations. Shim pulses
for spectroscopy are meant to be used intermittently during
stroboscopic FID readout; they work by correcting the phase of spins as
function of spatial position while leaving the chemical shift
evolution intact. Classical Hahn echo refocusing pulses are obviously
inadequate since they would refocus the chemical shift as well,
removing all spectral information. 

For imaging purposes, shim pulses can be used to impart the correct
phase at the center of k-space acquisition. The main challenges to
designing shim pulses are: 1) keeping the pulse as short as possible
so that large spectral widths can be used, 2) finding the best way to
impart a large enough phase modulation given the limited available
gradient strength during this short amount of time.

The adiabatic pulses of Topgaard are generally too long for many
applications and this limits the performance of many experiments. For
example, a 10 ms delay between consecutively collected samples results
in a maximum spectral width of 100 Hz, and readouts of several points
lead to significant $T_2$ relaxation.

Here, we exploit the idea that in an ex-situ NMR environment
(e.g. with a single-sided NMR magnet configuration), the static field
is intrinsically inhomogeneous and so is the radiofrequency (RF) field
due to the nature of single-sided magnet and coil designs. It thus
makes sense to think of schemes that take advantage of the inherent RF
inhomogeneity, available at no additional cost, to correct
for the negative effects of static field inhomogeneities.

I present a method which uses trains of hard pulses generated by
inhomogeneous RF fields. It is based on the approach of Meriles {\it
  et al.}~\cite{bib:pines_exsitu_science} in the sense that hard
pulses are used. However, we explicitly construct the desired
polynomials. In the limit of small flip angles, these
rotations commute and combine to create any desired polynomial in the
spatial variables $x$, $y$ and $z$. The calculations presented
herein demonstrate the ability of RF gradients to generate
$z$-rotation shim pulses or excitation pulses.

\section{Theory}

Consider the following product of four rotations:

\begin{equation}
 e^X e^Y e^{-X} e^{-Y} = e^{ [X,Y] + \dots }
\end{equation}

\noindent where the dots mean ``higher order terms''. In what follows,
we drop these higher order terms from the notation.

We observe that if $X$ contains the spin operator $I_x$ and $Y$
contains $I_y$, the commutator $[X,Y]$ will contain $I_z$. This
principle allows us to generate a rotation about the $z$
axis. Moreover, the product of $X$ and $Y$ in this commutator allows
us to generate polynomials in the coefficients of $X$ and $Y$.

We consider therefore rotation operators of the form:

\begin{equation}
 X_i = g_i x_i I_x, \qquad Y_j = h_j y_j I_y
\end{equation}

\noindent where $x_i$ and $y_i$ refer to any of the spatial variables
$x$, $y$ or $z$. The commutator of $X_j$ and $Y_j$ is:

\begin{equation}
 [X_j, Y_j] = i g_j h_j x_j y_j I_z
\end{equation}

\noindent and allows us to generate any of the following gradient
terms in the monomials $x^2$, $y^2$, $z^2$, $xy$, $yz$ or $xz$ by
picking $x_j$ and $y_j$ appropriately. Moreover, we may pick the
product $g_j h_j$ to produce any desired scaling (rotation angle).

\subsection{Second order shim pulse}

We may combine $N$ successive groups of 4 pulses back to back to
generate any linear combinations of $N$ monomials as follows:

\begin{equation}
 \prod_{j=1}^N \underbrace{e^{ X_j } e^{ Y_j } e^{ -X_j } e^{ -Y_j
 }}_{ \mbox{one monomial} } =
 \prod_{j=1}^N e^{ [ X_j, Y_j ] } = e^{ i I_z \sum_{j=1}^N g_j h_j x_j
 y_j }
\end{equation}

We will call this type of pulse a second order shim pulse (because it
contains the product $x_j y_j$ of spatial variables) and denote
the basic unit $e^{ X_j } e^{ Y_j } e^{ -X_j } e^{ -Y_j }$ as
$S_2(X_j,Y_j)$.

While this $S_2$ pulse creates an $I_z$ rotation, it is ideal for use
in stroboscopic pulse train experiments, where the phase of freely
evolving spins is periodically corrected. The $S_2$ pulse can be
converted into an excitation pulse using the following method:
the $S_2$ pulse is sandwiched on the left by a $(\pi/2)_y$ on the left
and $(\pi/2)^{-y}$ on the right. This has the effect of rotating $I_z$
into $I_x$, which can then serve as an excitation pulse. If multiple
subunits of $S_2$ are used, the sandwiching need only be applied to
the combined group of pulses. We note that the $(\pi/2)$ pulse
should be applied using a homogeneous RF field or using composite
pulses that compensate for the spatial inhomogeneity of the RF field.

\subsection{Third order shim pulse}

A third order shim pulse can be generated from the following sequence:

\begin{equation}
S_2(X,Y) e^{Z} \overline{ S_2(X,Y) } e^{-Z} = e^{ [ [X,Y],Z ] } \equiv
S_3(X,Y,Z)
\end{equation}

This double commutator is unable to generate an $I_z$ rotation if the
building blocks are the operators $I_x$ and $I_y$. In this case, the
$S_3$ pulse subunit generates an $I_x$ or $I_y$ rotation which can
serve as excitation pulse.

To get an $I_z$ rotation pulse, the following modification to $S_3$
can be used: an $I_x$ rotation can be converted into an $I_z$ rotation
using a $\pi/2$ rotation about $I_y$. Therefore, sandwiching the $S_3$
pulse, by  $(\pi/2)_y$ on the left and $(\pi/2)^{-y}$ on the right
converts it to an $I_z$ rotation. We note that the $(\pi/2)$ pulse
should be applied using a homogeneous RF field or using composite
pulses that compensate for the spatial inhomogeneity of the RF field.

\subsection{Fourth order shim pulse}

Generalizing these ideas, we may create a fourth order shim pulse can
using the following sequence:

\begin{multline}
 S_3(X,Y,Z) e^W S_3(X,Y,Z) e^{-W} = \\
 e^{ [[[X,Y],Z],W] } \equiv S_4(X,Y,Z,W)
\end{multline}

This sequence of commutators can generate an $I_z$ rotation, for
example, as follows:

\begin{equation}
 [[[ I_x, I_y ], I_x ] , I_x ] = - i I_z 
\end{equation}

\subsection{Higher order shim pulses}

A fifth order shim pulse is given by:

\begin{multline}
 S_4(X,Y,Z,W) e^{U} \overline{ X,Y,Z,W } e^{-U} = \\
 e^{ [[[[X,Y],Z],W],U] } \\
 \equiv S_5(X,Y,Z,W,U)
\end{multline}

\noindent whereas a sixth order shim pulse is the sequence:

\begin{multline}
 S_5(X,Y,Z,W,U) e^{V} \overline{ S_5(X,Y,Z,W,U) } e^{-V} = \\
 e^{ [[[[[X,Y],Z],W],U],V] } \\
 \equiv S_6(X,Y,Z,W,U,V)
\end{multline}

This latter pulse generates an $I_z$ rotation, for example, as
follows:

\begin{equation}
 [[[[[ I_x, I_y ], I_x], I_x], I_x], I_x] = - i I_z 
\end{equation}

{ This process of can be generalized to any $n$th order
monomial of the form $x_j^a y_j^b$ where $a+b=n$. By concatenating any
number of monomial units, we may generate any linear combinations of
them, by choosing the RF pulse amplitudes to match the desired
scalings of each monomial. This results in an arbitrary 
polynomial shim pulse. }

We note that the higher the order of the mononomial term, the higher
the error in the first order approximation to the rotation will
be. For example, a monomial such as $x^2$ grows rapidly with
increasing distance $x$ from the origin. Therefore, the errors grow
larger close the sample edges. One possible solution to this problem
is to repeat the $S_2$ sequence many times for smaller flip
angles. The error is then reduced to any desired order provided the
flip angles are made small enough. In the following section, I present
results that illustrate this refinement approach.

A second possible approach would be to take a basic sequence such as
$S_2$, repeat it a few times, and use optimal control algorithms (such
as GRAPE) to find the proper spatial variables, spin operators and
their relative phases in order to generate a propagator with far fewer
errors in the spatial and spin operator orders. The free parameters
which could be used in an optimal control optimization include pulse
phase, gradient direction ($x$, $y$ or $z$) and pulse
amplitude. Another method would be to use the inverse scattering
transform to solve for the RF waveform given the desired 
excitation profile. These methods will not be discussed here.

\begin{figure}[h]
\includegraphics[width=0.99\hsize]{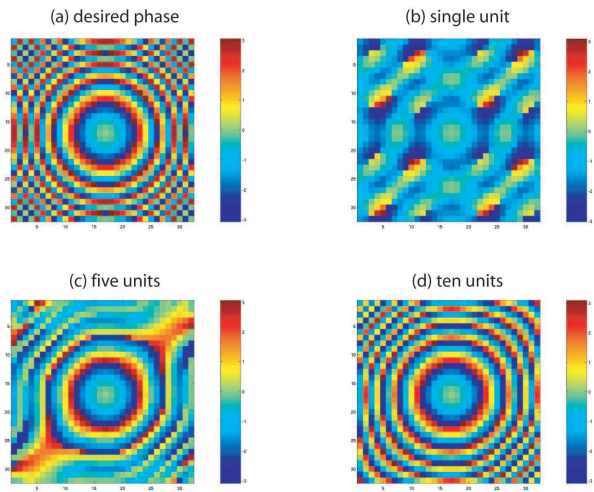}
 \caption{\label{fig:x2y2plots} Simulation of a second order $x^2+y^2$
 shim pulse: (a) the desired or ``target'' phase profile, (b) attempt
 at the shim pulse using a single unit of $S_2(X,Y)$, (c) five units
 and (d) ten units.}
\end{figure}

\section{The Refinement Method}

Figure~\ref{fig:x2y2plots} shows simulations of a second order shim
pulse $x^2+y^2$. Figure~\ref{fig:x2y2plots}(a) shows the ideal or
``target'' phase profile in the $xy$ plane. The pulse unit required
to generate this kind of shim pulse is

\begin{align}
S_2(X_1,Y_1) S_2(X_1,Y_1) =& e^{ x I_x } e^{ x I_y } e^{ -x I_x } e^{
  -x I_y } \nonumber \\
 &  \times e^{ y I_x } e^{ y I_y } e^{ -y I_x } e^{ -y I_y } \nonumber
\\
 =& e^{ i I_z (x^2 + y^2) }
\end{align}

This basic unit is the $N=2$ case. It requires a total of 8 hard
pulses. The phase obtained from a single unit of this shim pulse is
shown in Figure~\ref{fig:x2y2plots}(b). We see that the pulse performs
well at the origin but suffers from extreme distortions at large
values of $x^2+y^2$. This pulse can be concatenated into five
``smaller' (weaker amplitude) but otherwise identical RF pulses, as
shown in Figure~\ref{fig:x2y2plots}(c) with a slight improvement near
the center.

The ten units version of this pulse is shown in
Figure~\ref{fig:x2y2plots}(d) and reproduces well the desired (ideal)
phase pattern of Figure~\ref{fig:x2y2plots}(a). This pulse contains a
total of $8 \times 10 = 80$ hard pulses. If each hard pulses is 10
$\mu$s long, the total length of this pulse would be 800 $\mu$s. A 800
$\mu$s shim pulse is already an improvement by an order of magnitude
over the Topgaard adiabatic pulses.

In practice, the total duration of each hard pulse depends on the
available maximum $B_1$ amplitude. When considering the
time-optimality of this shim pulse, only the pulse area for each hard
pulse matters (duration $\times$ amplitude). Thus, in order to halve
the total duration of the shim pulse, we require the doubling of
$B_1$.

\section{Conclusion}

The basic idea of using RF gradients to generate arbitrary order shim
pulses works, as shown in the previous simulations and theoretical
expressions. Two basic errors arise in these analytical expressions:
for the generation of an $I_z$ shim pulse, the presence of $I_x$ and
$I_y$ which ``contaminate'' a desired $I_z$ rotation, and the actual
coefficient of $I_z$ contains not only the desired monomial but
monomials which are higher order in the spatial variables. A similar
argument applies to $I_x$ or $I_y$ pulses for excitation purposes.

It is likely that the performance of such pulses could be
improved by optimal control methods, where the spatial variables
$x_j$, $y_j$, $z_j$ pulse amplitudes $g_j, h_j$ and pulse phases $I_x$
vs. $I_y$ (or intermediate phases) are used as optimization parameters
during the search. This requires that some fidelity measure for the
propagator is used that minimizes the errors in the ``outer parts'' of
the spatial variables, where the pulse errors are largest.
By allowing arbitrary choices of the spatial variables and their
coefficient in the optimization algorithm, it could be possible to
eliminate the higher order spatial terms (monomials) in $I_z$ via
cancellations (negative terms could possibly cancel positive terms).
By allowing the phase of the RF pulse elements to vary, additional
desirable elements such as robustness to offsets, and elimination of
spurious spin operator terms $I_x$, $I_y$ from the propagator of a
desired $I_z$ rotation would be possible.


\bibliographystyle{unsrt}
\bibliography{dbase8}

\end{document}